\documentclass[12pt]{article}

\begin{document}

\title{\bf Matter Collineations of Plane Symmetric Spacetimes}

\author{M. Sharif \thanks{msharif@math.pu.edu.pk} and Nousheen Ilyas\\
Department of Mathematics, University of the Punjab,\\
Quaid-e-Azam Campus, Lahore-54590, Pakistan.}
\date{}

\maketitle

\begin{abstract}
This paper is devoted to the study of matter collineations of plane
symmetric spacetimes (for a particular class of spacetimes) when the
energy-momentum tensor is non-degenerate. There exists many
interesting cases where we obtain proper matter collineations. The
matter collineations in these cases are {\it four}, \emph{five},
{\it six}, \emph{seven} and {\it ten} with some constraints on the
energy-momentum tensor. We have solved some of these constraints to
obtain solutions of the Einstein field equations.
\end{abstract}

\date{}

\section{Introduction}

It has been an interesting subject to use the symmetry group of a
spacetime in constructing the solution of the Einstein field
equation (EFEs) given by
\begin{equation}
G_{ab}\equiv R_{ab}-\frac 12 Rg_{ab}=\kappa T_{ab},
\end{equation}
where $G_{ab}$ are components of the Einstein tensor, $R_{ab}$ are
components of the Ricci tensor and $T_{ab}$ are components of the
matter (energy-momentum) tensor, $R$ is the Ricci scalar and
$\kappa$ is the gravitational constant. Further, these solutions are
classified according to the Lie algebra structure generated by these
symmetries. Curvature and the Ricci tensors are the important
quantities which play a significant role in understanding the
geometric structure of spacetime. A pioneer study of curvature
collineations (CCs) and Ricci collineations (RCs) has been carried
out by Katzin, et al [1] and a further classification of CCs and RCs
has been obtained by different authors [2,3].

Symmetries of the Ricci tensor and, in particular, energy-momentum
tensor have recently been studied. The Einstien tensor quite
naturally arises in the theory of General Relativity and is related,
via EFEs, to the material content of the spacetime. Thus it has an
important role in this theory. In this paper, we shall analyze the
properties of a vector field along which the Lie derivative of the
energy-momentum tensor vanishes, i.e., ${\cal L}_\xi T_{ab}=0$.

A smooth vector field ${\bf \xi}$ is said to preserve a matter
symmetry [4] on $M$ if, for each smooth local diffeomorphism
$\phi_t$ associated with ${\bf \xi}$, the tensors $T$ and
$\phi^*_tT$ are equal on the domain $U$ of $\phi_t$, i.e.,
$T=\phi_t^*T$. Equivalently, a vector field $\xi^a$ is said to
generate a matter collineation if it satisfies the following
equation
\begin{equation}
\pounds_{\xi}T_{ab}=0,\quad  or\quad T_{ab,c} \xi^c + T_{ac}
\xi^c_{,b} + T_{cb} \xi^c_{,a}= 0,\quad(a,b,c=0,1,2,3),
\end{equation}
where $\pounds$ is the Lie derivative operator, $\xi^a$ is the
symmetry or collineation vector. Every Killing vector (KV) is a
matter collineation (MC) but the converse is not true, in general. A
proper MC is an MC which is not a KV, or a homothetic vector.

There is a growing interest in the study of MCs [5-7 and references
therein]. Hall, et al [5], in the discussion of RC and MC, have
argued that the symmetries of the energy-momentum tensor may also
provide some extra understanding of the subject which has not been
provided by KVs, Ricci and CCs. Carot, et al [6,7] have discussed
MCs from the point of view of the Lie algebra of vector fields
generating them and, in particular, he discussed spacetimes with a
degenerate $T_{ab}$. Some recent investigations [8-20] show keen
interest in the study of matter collineations. In the papers
[13-15], the study of MCs has been taken for spherically symmetric,
static plane symmetric and cylindrically symmetric spacetimes and
some interesting results have been obtained. Recently, the proper
MCs of non-static plane symmetric spacetimes have been found [19]
when the energy-momentum tensor is degenerate for spacetimes
satisfying $T_{01}=0$. It turns out that this admits an MC Lie
algebra of 4, 6, 10 dimensions apart from the infinite dimensional
algebras. This paper extends the above study to the non-degenerate
case.

The distribution of the paper is as follows. In the next section, we
write down the MC equations for plane symmetric spacetimes. In
section \textbf{3}, we calculate MCs by solving MC equations for the
non-degenerate case. Section \textbf{4} carries some examples
satisfying the constraints. Finally, a discussion of the results is
given in the last section.

\section{Matter Collineation Equations}

The most general form of the plane symmetric metric is given in the
form [21]
\begin{equation}
ds^2 =
e^{\nu(t,x)}dt^2-e^{\lambda(t,x)}dx^2-e^{\mu(t,x)}(dy^2+dz^2),
\end{equation}
where $\nu,~\lambda$ and $\mu$ are arbitrary functions of $t$ and
$x$. The surviving components of the energy-momentum tensor are
given as
\begin{eqnarray}
T_{00} &=&
\frac{1}{4}(\dot{\mu}^2+2\dot{\mu}\dot{\lambda})-\frac{1}{4}e^{v-\lambda}
(4\mu''+ 3 \mu'^2 - 2 \mu'\lambda'),\nonumber\\
T_{01}&=&-\frac{1}{2} (2 \dot{\mu}' + \dot{\mu}\mu' - \dot{\mu} v' -
\mu'\dot{ \lambda}), \nonumber \\
T_{11}&=&\frac{1}{4}({\mu'}^2+2\mu'\nu')-\frac{1}{4}e^{\lambda-\nu}
(4\ddot{\mu}+3\dot{\mu}^2-2\dot{\mu}\dot{\nu}),\nonumber\\
T_{22}& =&\frac{1}{4}e^{\mu-\lambda}(2\mu''+{\mu'}^2
-\mu'\lambda'+\mu'\nu'-\lambda'\nu'+{\nu'}^2+2\nu'')\nonumber\\
&-&\frac{1}{4}e^{\mu-\nu}(2\ddot{\mu}+\dot{\mu}^2
-\dot{\mu}\dot{\nu}+\dot{\mu}\dot{\lambda}-\dot{\nu}\dot{\lambda
}+\dot{\lambda}^2+2\ddot{\lambda}),\nonumber \\
T_{33}& =& T_{22}.
\end{eqnarray}
Here dot and prime indicate differentiation w.r.t. time and $x$
coordinate respectively. The MC equations can be written as follows
\begin{eqnarray}
2T_{01}\xi^1_{,0}+T_{00,0} \xi^0+T_{00,1} \xi^1
+ 2 T_{00} \xi^0_{,0} &=& 0,\\
T_{01,0}\xi^0+T_{01}\xi^0_{0}+T_{00} \xi^0_{,1}+T_{01,1}\xi^1
+T_{11} \xi^1_{,0}+T_{01}\xi^1_{,1} &=& 0,\\
T_{00} \xi^0_{,2}+T_{01}\xi^1_{,2}+T_{22} \xi^2_{,0} &=& 0,\\
T_{00} \xi^0_{,3}+T_{01}\xi^1_{,3}+T_{22} \xi^3_{,0} &=&0,\\
T_{11,0} \xi^0+2T_{01}\xi^0_{1}+T_{11,1} \xi^1 + 2 T_{11} \xi^1_{,1} &= &0,\\
T_{01}\xi^0_{,2}+T_{11} \xi^1_{,2}+T_{22} \xi^2_{,1} &= &0,\\
T_{01}\xi^0_{,3}+T_{11} \xi^1_{,3}+T_{22}\xi^3_{,1}&=&0,\\
T_{22,0} \xi^0+T_{22,1} \xi^1 + 2 T_{22}\xi^2_{,2}&=&0,\\
T_{22}(\xi^2_{,3}+\xi^3_{,2}) &=&0,\\
T_{22,0} \xi^0+T_{22,1} \xi^1+ 2T_{22}\xi^3_{,3}& =& 0.
\end{eqnarray}
These are the first order non-linear partial differential equations
in four variables $\xi^a(x^b)$. We restrict ourselves with the
assumption $T_{01}=0$ for the sake of simplicity. For this purpose,
either we need to make transformations such that $T_{01}=0$ or we
take such spacetimes which satisfy this condition. Here we take the
following class of metrics satisfying this assumption.
\begin{eqnarray*}
E(2))\otimes{\textbf{R}},~\mathrm{where}~{\textbf{R}}=\partial_x~
\mathrm{if~and~only~if}~\nu=\nu(t),~\lambda=\lambda(t),~\mu=2\ln t.
\end{eqnarray*}
For the sake of simplicity, we use the notation $T_{aa}=T_a$ when
$T_{01}=0$. The solution of these equations for the degenerate case,
i.e., $\det(T_{ab})=0$ for the above class of metrics has already
been completed [19]. In this paper, we are giving the complete
classification for the non-degenerate case with $T_{01}=0$.

\section{Solution for the Non-Degenerate Case}

This section explicitly provides a complete classification of plane
symmetric spacetimes for the non-degenerate case for the above
mentioned case.

For this case, MC Eqs.(5)-(14) reduce to
\begin{eqnarray}
T_{0,0}\xi^0+2T_{0}\xi^0_{,0}=0,\\
T_{0}\xi^0_{,1}+T_{1}\xi^1_{,0}=0,\\
T_{0}\xi^0_{,2}+T_{2}\xi^2_{,0}=0,\\
T_{0}\xi^0_{,3}+T_{2}\xi^3_{,0}=0,\\
T_{1,0}\xi^0+2T_{1}\xi^1_{,1}=0,\\
T_{1}\xi^1_{,2}+T_{2}\xi^2_{,1}=0,\\
T_{1}\xi^1_{,3}+T_{2}\xi^3_{,1}=0,\\
T_{2,0}\xi^0+2T_{2}\xi^2_{,2}=0,\\
T_{2}(\xi^2_{,3}+\xi^3_{,2})=0,\\
T_{2,0}\xi^0+2T_{2}\xi^3_{,3}=0.
\end{eqnarray}
When we solve Eqs.(15)-(24) simultaneously, after some algebraic
computations, we obtain
\begin{eqnarray}
\xi^0&=&-\frac{T_{2}}{T_{0}}\{\frac{1}{2}\dot{A_1}(y^2+z^2)
+\dot{A_2}z+\dot{A_3}y\}+A_4,\nonumber\\
\xi^1&=&-\frac{T_{2}}{T_{1}}\{\frac{1}{2}A'_1(y^2+z^2)+A'_2z+A'_3y\}+A_5,\nonumber\\
\xi^2&=&c_2yz+\frac{c_3}{2}(z^2-y^2)+c_4z+A_1y+A_3,\nonumber\\
\xi^3&=&\frac{c_2}{2}(z^2-y^2)-(c_3z+c_4)y+A_1z+A_2
\end{eqnarray}
subject to the constraints
\begin{eqnarray}
\dot{T}_{2}\dot{A}_1&=&0,\\
\frac{\dot{T}_{1}}{T_{0}}\dot{A}_i+2A''_i&=&0,\\
\frac{\dot{T}_{2}}{T_{2}}A_4+2A_1&=&0,\\
\frac{\dot{T}_{2}}{T_{2}\sqrt{T_{0}}}g_2(x)-2c_2&=&0,\\
\frac{\dot{T}_{2}}{T_{2}\sqrt{T_{0}}}g_3(x)+2c_3&=&0,\\
\frac{\dot{T}_{2}}{T_{2}\sqrt{T_{0}}}g_4(x)+2A_1&=&0,\\
\frac{\dot{T}_{1}}{T_{2}\sqrt{T_{0}}}g_i(x)+2A''_i&=&0,\\
A'_i=\sqrt{\frac{T_{1}}{T_{2}}}f_i(x),\quad
\dot{A}_i&=&\frac{\sqrt{T_{0}}}{T_{2}}g_i(x),\quad(i=1,2,3)\\
A'_5=-\frac{\dot{T_1}}{2T_{1}\sqrt{T_{0}}}g_4(x), \quad
\dot{A}_5&=&-\frac{\sqrt{T_{0}}}{T_{1}}g'_4(x),
\end{eqnarray}
where $c_2,~ c_3,~ c_4$ are arbitrary constants and
$A_\mu=A_\mu(t,x)(\mu=1,2,3,4,5)$, $f_i(x),~ g_i(x),~ g_4(x)$ are
integration functions. Thus the problem is reduced to solving the
set of Eqs.(15)-(24) subject to the constraint Eqs.(26)-(34). From
Eqs.(29) and (30), there arise the following two cases:
\begin{eqnarray*}
(1)\quad(\frac{\dot{T}_{2}}{T_{2}\sqrt{T_{0}}})^.\neq0,\quad
(2)\quad(\frac{\dot{T}_{2}}{T_{2}\sqrt{T_{0}}})^.=0.
\end{eqnarray*}
\textbf{Case (1):} In this case, we have $\dot{T}_{2}\neq0$ and
Eq.(26) yields $A_1= A_1(x)$. Then it follows from Eq.(31) that
\begin{equation}
\frac{\dot{T}_{2}}{T_{2}\sqrt{T_{0}}}g_4(x)+2A_1(x)=0
\end{equation}
which is possible only if $g_4=0=A_1$. Also, from Eqs.(28) and
(34), we have $A_4=0,~ A_5=c_0$. Now Eqs.(29) and (30) yield that
$g_2=0=g_3,~ c_2=0=c_3$ and from Eqs.(32) and (33), we have
$f_j(x)=a_j,~(j=2,3)$, where $a_j$ are arbitrary constants. Also,
it follows from Eq.(33) that
\begin{eqnarray}
\dot{A}_j=0,\quad A'_j=\sqrt{\frac{T_{1}}{T_{2}}}a_j\nonumber
\end{eqnarray}
which gives
\begin{eqnarray}
A_j&=&\sqrt{\frac{T_{1}}{T_{2}}}a_jx+h_j(t),\\
(\sqrt{\frac{T_{1}}{T_{2}}})^.a_jx+\dot{h}_j(t)&=&0.\nonumber
\end{eqnarray}
This implies that either
\begin{eqnarray*}
(i)\quad(\sqrt{\frac{T_{1}}{T_{2}}})^.=0,\quad \mathrm{or}\quad
(ii)\quad(\sqrt{\frac{T_{1}}{T_{2}}})^.\neq0.
\end{eqnarray*}
For both the cases $h_j(t)=b_j$, where $b_j, (j=2,3)$ are
arbitrary constants and hence Eq.(36) takes the form
\begin{equation}
A_j=\sqrt{\frac{T_{1}}{T_{2}}}a_jx+b_j
\end{equation}
\textbf{Case 1(i):} Here we have $(\frac{T_{1}}{T_{2}})^.=0$ which
implies that $T_{1}=dT_{2}$, where $d$ is an arbitrary constant.
Thus Eq.(37) reduces to
\begin{equation}
A_j=d_jx+b_j,\quad(j=2,3)
\end{equation}
where $d_j~(j=2,3)$ are arbitrary constants. Consequently, we obtain
six independent MCs out of which three are the minimal KVs of plane
symmetric and three are proper MCs. These are given by
\begin{eqnarray}
\xi_{(1)}&=&\partial_y,\quad
\xi_{(2)}=\partial_z,\quad\xi_{(3)}=z\partial_y-y\partial_z,\nonumber\\
\xi_{(4)}&=&\partial_x,\quad\xi_{(5)}=x\partial_z-\frac{T_{2}}{T_{1}}z\partial_x,\quad
\xi_{(6)}=x\partial_y-\frac{T_{2}}{T_{1}}y\partial_x.
\end{eqnarray}
The corresponding Lie algebra is given by
\begin{eqnarray*}
[\xi_{(1)},\xi_{(3)}]=-\xi_{(2)},\quad
[\xi_{(1)},\xi_{(6)}]=-\frac{T_2}{T_1}\xi_{(4)},\quad
[\xi_{(2)},\xi_{(3)}]=\xi_{(1)},
\end{eqnarray*}
\begin{eqnarray*}
[\xi_{(2)},\xi_{(5)}]=-\frac{T_{2}}{T_{1}}\xi_{(4)},\quad
[\xi_{(3)},\xi_{(5)}]&=&\xi_{(6)},\quad
[\xi_{(3)},\xi_{(6)}]=-\xi_{(5)},
\end{eqnarray*}
\begin{eqnarray*}
[\xi_{(4)},\xi_{(5)}]=\xi_{(2)},\quad[\xi_{(4)},\xi_{(6)}]=\xi_{(1)}
,\quad [\xi_{(5)},\xi_{(6)}]=-\frac{T_{2}}{T_{1}}\xi_{(3)},
\end{eqnarray*}
\begin{eqnarray*}
[\xi_{(i)},\xi_{(j)}]=0,\quad \mathrm{(otherwise)}.
\end{eqnarray*}
\textbf{Case 1(ii):} Here $(\sqrt{\frac{T_{1}}{T_{2}}})^.\neq0$
which implies that $a_j=0$ and then Eq.(37) reduces to $A_j=b_j$.
Solving the MCs equations, we obtain three usual plane symmetry
KVs and one proper MC.\\
\textbf{Case 2.} This case yields
$\quad(\frac{\dot{T}_{2}}{T_{2}\sqrt{T_{0}}})=\alpha$, where
$\alpha$ is an arbitrary constant which can be zero or non-zero.\\
\textbf{Case 2(i):} First we take $\alpha\neq0$ which implies
$\dot{T}_{2}\neq0$. It follows from Eqs.(26), (29) and (30) that
\begin{equation}
g_2=\frac{2c_2}{\alpha},\quad g_3=-\frac{2c_3}{\alpha},\quad
A_1=A_1(x)
\end{equation}
and hence Eq.(33) yields
\begin{equation}
(\frac{T_{2}}{T_{1}})^.A'_i=0.
\end{equation}
This implies further two possibilities:
\begin{eqnarray*}
(a)\quad(\frac{T_{2}}{T_{1}})^.\neq0,\quad(b)\quad(\frac{T_{2}}{T_{1}})^.=0.
\end{eqnarray*}
For the \textbf{case 2i(a)}, Eq.(40) implies that $A_1=c_5$ and
consequently Eqs.(27), (28) and (34) respectively yield
\begin{eqnarray}
\dot{T}_{1}\dot{A}_i=0, \quad A_4=-\frac{2c_5}{\alpha\sqrt{T_{0}}},\\
\dot{A}_5=0,\quad A'_5=\frac{{\dot{T}_{1}}}{\alpha{T_{1}}\sqrt{T_{0}}}c_5,\\
A_5=\frac{{\dot{T}_{1}}}{\alpha{T_{1}}\sqrt{T_{0}}}c_5x+c_0.
\end{eqnarray}
Eq.(43) further gives two possibilities
\begin{eqnarray*}
(*)\quad(\frac{{\dot{T}_{1}}}{{T_{1}}\sqrt{T_{0}}})^.\neq0,
\quad(**)\quad(\frac{{\dot{T}_{1}}}{{T_{1}}\sqrt{T_{0}}})^.=0.
\end{eqnarray*}
In the \textbf{case 2ia$(*)$}, $\dot{T}_{1}\neq0$, we obtain three
usual plane symmetry KVs and one proper MC.

For the \textbf{case 2ia$(**)$}, we have
$\frac{{\dot{T}_{1}}}{{T_{1}}\sqrt{T_{0}}}=\beta$, where $\beta$
is an arbitrary constant which can be zero or non-zero. When
$\beta\neq0$, the proper MCs are given by
\begin{eqnarray}
\xi_{(4)}=\partial_x,\quad
\xi_{(5)}=-\frac{2}{\alpha\sqrt{T_{0}}}\partial_t+\frac{\beta}{\alpha}x\partial_x
+y\partial_y+z\partial_z.
\end{eqnarray}
The Lie algebra will become
\begin{eqnarray*}
[\xi_{(1)},\xi_{(3)}]=-\xi_{(2)},\quad
[\xi_{(1)},\xi_{(5)}]=\xi_{(1)},\quad
[\xi_{(2)},\xi_{(3)}]=\xi_{(1)},
\end{eqnarray*}
\begin{eqnarray*}
[\xi_{(2)},\xi_{(5)}]=\xi_{(2)},\quad
[\xi_{(4)},\xi_{(5)}]=\frac{\beta}{\alpha}\xi_{(4)},\quad
[\xi_{(i)},\xi_{(j)}]=0,\quad \mathrm{(otherwise)}.
\end{eqnarray*}
For $\beta=0$, the proper MCs are
\begin{eqnarray}
\xi_{(4)}&=&\partial_x,\nonumber\\
\xi_{(5)}&=&-\frac{2}{\alpha\sqrt{T_{0}}}z\partial_t
+yz\partial_y+\{\frac{1}{2}(z^2-y^2)
-\frac{2}{\alpha^2T_{2}}\}\partial_z,\nonumber\\
\xi_{(6)}&=&\frac{2}{\alpha\sqrt{T_{0}}}y\partial_t
+\{\frac{1}{2}(z^2-y^2)+\frac{2}{\alpha^2
T_{2}}\}\partial_y-yz\partial_z,\nonumber\\
\xi_{(7)}&=&-\frac{2}{\alpha\sqrt{T_{0}}}\partial_t+y\partial_y+z\partial_z.
\end{eqnarray}
The Lie algebra is
\begin{eqnarray*}
[\xi_{(1)},\xi_{(3)}]=-\xi_{(2)},\quad[\xi_{(1)},\xi_{(5)}]=\xi_{(3)},\quad
[\xi_{(1)},\xi_{(6)}]=-\xi_{(7)},
\end{eqnarray*}
\begin{eqnarray*}
[\xi_{(1)},\xi_{(7)}]=\xi_{(1)},\quad[\xi_{(2)},\xi_{(3)}]=\xi_{(1)},\quad
[\xi_{(2)},\xi_{(5)}]=\xi_{(7)},
\end{eqnarray*}
\begin{eqnarray*}
[\xi_{(2)},\xi_{(6)}]=\xi_{(3)},\quad[\xi_{(2)},\xi_{(7)}]=\xi_{(2)},\quad
[\xi_{(3)},\xi_{(5)}]=\xi_{(6)},
\end{eqnarray*}
\begin{eqnarray*}
[\xi_{(3)},\xi_{(6)}]=-\xi_{(5)},
\quad[\xi_{(5)},\xi_{(7)}]=-\xi_{(5)},\quad
[\xi_{(6)},\xi_{(7)}]=-\xi_{(6)},
\end{eqnarray*}
\begin{eqnarray*}
[\xi_{(i)},\xi_{(j)}]=0, \quad\mathrm{(otherwise)}.
\end{eqnarray*}

For the \textbf{case 2i(b)}, $T_{2}=\gamma T_{1}$, where $\gamma$
is non-zero arbitrary constant, the proper MCs are
\begin{eqnarray}
\xi_{(4)}&=&\partial_x,\nonumber\\
\xi_{(5)}&=&\frac{x}{\sqrt{\gamma}}\partial_z-\sqrt{\gamma}z\partial_x,\nonumber\\
\xi_{(6)}&=&-\frac{2}{\alpha
\sqrt{T_{0}}}z\partial_t+zx\partial_x+yz\partial_y
+\frac{1}{2}(z^2-y^2-\frac{x^2}{\gamma}-\frac{4}{\alpha^2 T_{2}})\partial_z,\nonumber\\
\xi_{(7)}&=&\frac{2}{\alpha
\sqrt{T_{0}}}y\partial_t-yx\partial_x+\frac{1}{2}(z^2-y^2
+\frac{x^2}{\gamma}+\frac{4}{\alpha^2 T_{2}})\partial_y-yz\partial_z,\nonumber\\
\xi_{(8)}&=&-\frac{2}{\alpha
\sqrt{T_{0}}}x\partial_t+\frac{\gamma}{2}(\frac{x^2}{\gamma}-y^2-z^2-\frac{4}{\alpha^2
T_{2}})\partial_x
+xy\partial_y+xz\partial_z,\nonumber\\
\xi_{(9)}&=&-\frac{2}{\alpha \sqrt{T_{0}}}\partial_t+x\partial_x
+y\partial_y+z\partial_z,\nonumber\\
\xi_{(10)}&=&\frac{x}{\sqrt{\gamma}}\partial_y-\sqrt{\gamma}y\partial_x.
\end{eqnarray}
The Lie algebra is given by
\begin{eqnarray*}
[\xi_{(1)},\xi_{(3)}]=-\xi_{(2)},\quad
[\xi_{(1)},\xi_{(6)}]=\xi_{(3)},\quad
[\xi_{(1)},\xi_{(7)}]=-\xi_{(9)},
\end{eqnarray*}
\begin{eqnarray*}
[\xi_{(1)},\xi_{(8)}]=\sqrt{\gamma}\xi_{(10)},\quad
[\xi_{(1)},\xi_{(9)}]=\xi_{(1)},\quad
[\xi_{(1)},\xi_{(10)}]=-\sqrt{\gamma}\xi_{(4)},
\end{eqnarray*}
\begin{eqnarray*}
[\xi_{(2)},\xi_{(3)}]=\xi_{(1)},\quad
[\xi_{(2)},\xi_{(5)}]=-\sqrt{\gamma}\xi_{(4)},\quad
[\xi_{(2)},\xi_{(6)}]=\xi_{(9)},
\end{eqnarray*}
\begin{eqnarray*}
[\xi_{(2)},\xi_{(7)}]=\xi_{(3)},\quad
[\xi_{(2)},\xi_{(8)}]=\sqrt{\gamma}\xi_{(5)},\quad
[\xi_{(2)},\xi_{(9)}]=\xi_{(2)},
\end{eqnarray*}
\begin{eqnarray*}
[\xi_{(3)},\xi_{(5)}]=-\xi_{(10)},\quad
[\xi_{(3)},\xi_{(6)}]=\xi_{(7)},\quad
[\xi_{(3)},\xi_{(7)}]=-\xi_{(6)},
\end{eqnarray*}
\begin{eqnarray*}
[\xi_{(3)},\xi_{(10)}]=\xi_{(5)},\quad
[\xi_{(4)},\xi_{(5)}]=\frac{1}{\sqrt{\gamma}}\xi_{(2)},\quad
[\xi_{(4)},\xi_{(6)}]=-\frac{1}{\sqrt{\gamma}}\xi_{(5)},
\end{eqnarray*}
\begin{eqnarray*}
[\xi_{(4)},\xi_{(7)}]=\frac{1}{\sqrt{\gamma}}\xi_{(10)},\quad
[\xi_{(4)},\xi_{(8)}]=\xi_{(9)},\quad
[\xi_{(4)},\xi_{(9)}]=\xi_{(4)},
\end{eqnarray*}
\begin{eqnarray*}
[\xi_{(4)},\xi_{(10)}]=\frac{1}{\sqrt{\gamma}}\xi_{(1)},\quad
[\xi_{(5)},\xi_{(6)}]=\frac{1}{\sqrt{\gamma}}\xi_{(8)},\quad
[\xi_{(5)},\xi_{(8)}]=-\sqrt{\gamma}\xi_{(6)},
\end{eqnarray*}
\begin{eqnarray*}
[\xi_{(5)},\xi_{(10)}]=-\xi_{(3)},\quad
[\xi_{(6)},\xi_{(9)}]=-\xi_{(6)},\quad
[\xi_{(7)},\xi_{(9)}]=-\xi_{(7)},
\end{eqnarray*}
\begin{eqnarray*}
[\xi_{(7)},\xi_{(10)}]=\frac{1}{\sqrt{\gamma}}\xi_{(8)},\quad
[\xi_{(8)},\xi_{(9)}]=-\xi_{(8)},\quad
[\xi_{(8)},\xi_{(10)}]=-\sqrt{\gamma}\xi_{(7)},
\end{eqnarray*}
\begin{eqnarray*}
[\xi_{(i)},\xi_{(j)}]=0,\quad \mathrm{(otherwise)}.
\end{eqnarray*}
\textbf{Case 2(ii):} Here $\alpha=0$ and we have $T_{2}=constant$.
Eqs.(29), (30) and (31) yield $c_2=0=c_3,~ A_1=0$ and Eq.(33) gives
\begin{equation}
(\frac{(\sqrt{T_{1}})^.}{\sqrt{T_{0}}})^.f_j(x)=0.
\end{equation}
This leads to two possibilities
\begin{eqnarray*}
(a)\quad(\frac{(\sqrt{T_{1}})^.}{\sqrt{T_{0}}})^.=0,\quad
(b)\quad(\frac{(\sqrt{T_{1}})^.}{\sqrt{T_{0}}})^.\neq0.
\end{eqnarray*}
For the \textbf{case 2ii(a)}, we have
$\frac{(\sqrt{T_{1}})^.}{\sqrt{T_{0}}}=\delta$, where $\delta$ is
an arbitrary constant which can be zero or non-zero.

In the \textbf{case 2iia$(*$)}, when $\delta\neq0$, the proper MCs
are
\begin{eqnarray}
\xi_{(4)}&=&\partial_x,\nonumber\\
\xi_{(5)}&=&\frac{1}{\sqrt{T_{0}}}\cos\delta
x\partial_t-\frac{1}{\sqrt{T_{1}}}\sin\delta
x\partial_x,\nonumber\\
\xi_{(6)}&=&\frac{1}{\sqrt{T_{0}}}\sin\delta
x\partial_t+\frac{1}{\sqrt{T_{1}}}\cos\delta
x\partial_x,\nonumber\\
\xi_{(7)}&=&-\frac{z}{\sqrt{T_{0}}}\cos\delta
x\partial_t+\frac{z}{\sqrt{T_{1}}}\sin\delta
x\partial_x+\frac{\sqrt{T_{1}}}{\delta T_{2}}\cos\delta
x\partial_z,\nonumber\\
\xi_{(8)}&=&-\frac{z}{\sqrt{T_{0}}}\sin\delta
x\partial_t-\frac{z}{\sqrt{T_{1}}}\cos\delta
x\partial_x+\frac{\sqrt{T_{1}}}{\delta T_{2}}\sin\delta
x\partial_z,\nonumber\\
\xi_{(9)}&=&-\frac{y}{\sqrt{T_{0}}}\cos\delta
x\partial_t+\frac{y}{\sqrt{T_{1}}}\sin\delta
x\partial_x+\frac{\sqrt{T_{1}}}{\delta T_{2}}\cos\delta
x\partial_y,\nonumber\\
\xi_{(10)}&=&-\frac{y}{\sqrt{T_{0}}}\sin\delta
x\partial_t-\frac{y}{\sqrt{T_{1}}}\cos\delta
x\partial_x+\frac{\sqrt{T_{1}}}{\delta T_{2}}\sin\delta
x\partial_y.
\end{eqnarray}
The Lie algebra is given by
\begin{eqnarray*}
[\xi_{(1)},\xi_{(3)}]=-\xi_{(2)},\quad
[\xi_{(1)},\xi_{(9)}]=-\xi_{(5)},\quad
[\xi_{(1)},\xi_{(10)}]=-\xi_{(6)},
\end{eqnarray*}
\begin{eqnarray*}
[\xi_{(2)},\xi_{(3)}]=\xi_{(1)},\quad
[\xi_{(2)},\xi_{(7)}]=-\xi_{(5)},\quad
[\xi_{(2)},\xi_{(8)}]=-\xi_{(6)},
\end{eqnarray*}
\begin{eqnarray*}
[\xi_{(3)},\xi_{(7)}]=-\xi_{(9)},\quad
[\xi_{(3)},\xi_{(8)}]=-\xi_{(10)},\quad
[\xi_{(3)},\xi_{(9)}]=\xi_{(7)},
\end{eqnarray*}
\begin{eqnarray*}
[\xi_{(3)},\xi_{(10)}]=\xi_{(8)},\quad
[\xi_{(4)},\xi_{(5)}]=-\delta\xi_{(6)},\quad
[\xi_{(4)},\xi_{(6)}]=\delta\xi_{(5)},
\end{eqnarray*}
\begin{eqnarray*}
[\xi_{(4)},\xi_{(7)}]=-\delta\xi_{(8)},\quad
[\xi_{(4)},\xi_{(8)}]=\delta\xi_{(7)},\quad
[\xi_{(4)},\xi_{(9)}]=-\delta\xi_{(10)},
\end{eqnarray*}
\begin{eqnarray*}
[\xi_{(4)},\xi_{(10)}]=\delta\xi_{(9)},\quad
[\xi_{(5)},\xi_{(7)}]=\frac{1}{T_{2}}\xi_{(2)},\quad
[\xi_{(5)},\xi_{(9)}]=\frac{1}{T_{2}}\xi_{(1)},
\end{eqnarray*}
\begin{eqnarray*}
[\xi_{(6)},\xi_{(8)}]=\frac{1}{T_{2}}\xi_{(2)},\quad
[\xi_{(6)},\xi_{(10)}]=\frac{1}{T_{2}}\xi_{(1)},\quad
[\xi_{(7)},\xi_{(9)}]=-\frac{1}{T_{2}}\xi_{(3)},
\end{eqnarray*}
\begin{eqnarray*}
[\xi_{(8)},\xi_{(10)}]=\frac{1}{T_{2}}y\xi_{(2)},\quad
[\xi_{(9)},\xi_{(10)}]=-\frac{1}{T_{2}}y\xi_{(1)},
\end{eqnarray*}
\begin{eqnarray*}
[\xi_{(i)},\xi_{(j)}]=0,\quad\mathrm{(otherwise)}.
\end{eqnarray*}
For the \textbf{case 2iia$(**$)}, i.e., $\delta=0$, $T_1=constant$
and the proper MCs are given by
\begin{eqnarray}
\xi_{(4)}&=&\partial_x,\nonumber\\
\xi_{(5)}&=&-\frac{T_{2}}{T_{1}}z\partial_x
+x\partial_z,\nonumber\\
\xi_{(6)}&=&-\frac{T_{2}}{T_{1}}y\partial_x+x\partial_y,\nonumber\\
\xi_{(7)}&=&-\frac{T_{2}}{\sqrt{T_{0}}}z\partial_t
+\int\sqrt{T_{0}}dt\partial_z,\nonumber\\
\xi_{(8)}&=&-\frac{T_{2}}{\sqrt{T_{0}}}y\partial_t
+\int\sqrt{T_{0}}dt\partial_y,\nonumber\\
\xi_{(9)}&=&\frac{1}{\sqrt{T_{0}}}\partial_t,\nonumber\\
\xi_{(10)}&=&\frac{x}{\sqrt{T_{0}}}\partial_t
-\frac{1}{T_{1}}\int\sqrt{T_{0}}dt\partial_x.
\end{eqnarray}
The Lie algebra is
\begin{eqnarray*}
[\xi_{(1)},\xi_{(3)}]=-\xi_{(2)},\quad
[\xi_{(1)},\xi_{(6)}]=-\frac{T_{2}}{T_{1}}\xi_{(4)},\quad
[\xi_{(1)},\xi_{(8)}]=-T_{2}\xi_{(9)},
\end{eqnarray*}
\begin{eqnarray*}
[\xi_{(2)},\xi_{(3)}]=\xi_{(1)},\quad
[\xi_{(2)},\xi_{(5)}]=-\frac{T_{2}}{T_{1}}\xi_{(4)},\quad
[\xi_{(2)},\xi_{(7)}]=-T_{2}\xi_{(9)},
\end{eqnarray*}
\begin{eqnarray*}
[\xi_{(3)},\xi_{(5)}]=-\xi_{(6)},\quad
[\xi_{(3)},\xi_{(6)}]=\xi_{(5)},\quad
[\xi_{(3)},\xi_{(7)}]=-\xi_{(8)},
\end{eqnarray*}
\begin{eqnarray*}
[\xi_{(3)},\xi_{(8)}]=\xi_{(7)},\quad
[\xi_{(4)},\xi_{(5)}]=\xi_{(2)},\quad
[\xi_{(4)},\xi_{(6)}]=\xi_{(1)},
\end{eqnarray*}
\begin{eqnarray*}
[\xi_{(4)},\xi_{(10)}]=\xi_{(9)},\quad
[\xi_{(5)},\xi_{(6)}]=-\frac{T_{2}}{T_{1}}\xi_{(3)},\quad
[\xi_{(5)},\xi_{(7)}]=-T_{2}\xi_{(10)},
\end{eqnarray*}
\begin{eqnarray*}
[\xi_{(5)},\xi_{(10)}]=\frac{1}{T_{1}}\xi_{(7)},\quad
[\xi_{(6)},\xi_{(8)}]=-T_{2}\xi_{(10)},\quad
[\xi_{(6)},\xi_{(10)}]=\frac{1}{T_{1}}\xi_{(8)},
\end{eqnarray*}
\begin{eqnarray*}
[\xi_{(7)},\xi_{(8)}]=-T_{2}\xi_{(3)},\quad
[\xi_{(7)},\xi_{(10)}]=-\xi_{(5)},\quad
[\xi_{(8)},\xi_{(10)}]=-\xi_{(6)},
\end{eqnarray*}
\begin{eqnarray*}
[\xi_{(9)},\xi_{(10)}]=-\frac{1}{T_{1}}\xi_{(4)},\quad
[\xi_{(i)},\xi_{(j)}]=0,\quad\mathrm{(otherwise)}.
\end{eqnarray*}
For the \textbf{case 2ii(b)}, Eqs.(27) and (33) yield $A_j=a_j$ and
hence Eq.(34) gives
\begin{equation}
g''_4(x)=\frac{T_{1}}{2\sqrt{T_{0}}}(\frac{\dot{T}_{1}}{T_{1}\sqrt{T_{0}}})^.g_4(x).
\end{equation}
This leads to the following two possibilities
\begin{eqnarray*}
(*)\quad(\frac{T_{1}}{2\sqrt{T_{0}}}(\frac{\dot{T}_{1}}{T_{1}\sqrt{T_{0}}})^.)^.=0,\quad
(**)\quad(\frac{T_{1}}{2\sqrt{T_{0}}}(\frac{\dot{T}_{1}}{T_{1}\sqrt{T_{0}}})^.)^.\neq0.
\end{eqnarray*}
In the \textbf{case 2iib$(*$)}, we have
$\frac{T_{1}}{2\sqrt{T_{0}}}(\frac{\dot{T}_{1}}{T_{1}\sqrt{T_{0}}})^.=\eta$,
where $\eta$ is an integration constant which gives further three
cases according as $\eta>0$, or $\eta=0$ or $\eta<0$.

When $\eta=0$, i.e., \textbf{case 2iib$*(+)$}, we have
$\frac{\dot{T}_{1}}{T_{1}\sqrt{T_{0}}}=\epsilon$, where $\epsilon$
is non-zero arbitrary constant, the proper MCs are
\begin{eqnarray}
\xi_{(4)}&=&\partial_x,\nonumber\\
\xi_{(5)}&=&\frac{x}{\sqrt{T_{0}}}\partial_t+(\frac{1}{\epsilon
T_{1}}-\frac{\epsilon}{4}x^2)\partial_x,\nonumber\\
\xi_{(6)}&=&\frac{1}{\sqrt{T_{0}}}\partial_t-\frac{\epsilon}{2}x\partial_x.
\end{eqnarray}
The Lie algebra is given by
\begin{eqnarray*}
[\xi_{(1)},\xi_{(3)}]=-\xi_{(2)},\quad
[\xi_{(2)},\xi_{(3)}]=\xi_{(1)},\quad
[\xi_{(4)},\xi_{(5)}]=\xi_{(6)},
\end{eqnarray*}
\begin{eqnarray*}
[\xi_{(4)},\xi_{(6)}]=-\frac{\epsilon}{2}~\xi_{(4)},\quad
[\xi_{(5)},\xi_{(6)}]=\frac{\epsilon}{2}\xi_{(5)},\quad
[\xi_{(i)},\xi_{(j)}]=0,\quad\mathrm{(otherwise)}.
\end{eqnarray*}
For the \textbf{case 2iib$*(++)$},  we have the following proper MCs
\begin{eqnarray}
\xi_{(4)}&=&\partial_x,\nonumber\\
\xi_{(5)}&=&\frac{\cosh\sqrt{\eta}x}{\sqrt{T_{0}}}\partial_t
-(\frac{\dot{T}_{1}}{2T_{1}\sqrt{T_{0}}})
\frac{\sinh\sqrt{\eta}x}{\sqrt{\eta}}\partial_x,\nonumber\\
\xi_{(6)}&=&\frac{\sinh\sqrt{\eta}x}{\sqrt{T_{0}}}\partial_t
-(\frac{\dot{T}_{1}}{2T_{1}\sqrt{T_{0}}})
\frac{\cosh\sqrt{\eta}x}{\sqrt{\eta}}\partial_x
\end{eqnarray}
and the corresponding Lie algebra is
\begin{eqnarray*}
[\xi_{(1)},\xi_{(3)}]=-\xi_{(2)},\quad
[\xi_{(2)},\xi_{(3)}]=\xi_{(1)},\quad
[\xi_{(4)},\xi_{(5)}]=\sqrt{\eta}\xi_{(6)},
\end{eqnarray*}
\begin{eqnarray*}
[\xi_{(4)},\xi_{(6)}]=\sqrt{\eta}\xi_{(5)},\quad
[\xi_{(5)},\xi_{(6)}]=-\frac{1}{\sqrt{\eta}}\{\frac{\eta}{T_{1}}
+(\frac{\dot{T}_{1}}{2T_{1}\sqrt{T_{0}}})^2\}\xi_{(4)},
\end{eqnarray*}
\begin{eqnarray*}
[\xi_{(i)},\xi_{(j)}]=0,\quad\mathrm{(otherwise)}.
\end{eqnarray*}
In the \textbf{case 2iib$*(+++)$}, we take take $\psi=-\eta$, where
$\psi$ is also an arbitrary constant and the proper MCs turn out to
be
\begin{eqnarray}
\xi_{(4)}&=&\partial_x,\nonumber\\
\xi_{(5)}&=&\frac{\cos\sqrt{\psi}x}{T_{0}}\partial_t+(\frac{\dot{T}_{1}}{2T_{1}\sqrt{T_{0}}})
\frac{\sin\sqrt{\psi}x}{\sqrt{\psi}}\partial_x
,\nonumber\\
\xi_{(6)}&=&\frac{\sin\sqrt{\psi}x}{T_{0}}\partial_t-(\frac{\dot{T}_{1}}{2T_{1}\sqrt{T_{0}}})
\frac{\cos\sqrt{\psi}x}{\sqrt{\psi}}\partial_x .
\end{eqnarray}
The Lie algebra is
\begin{eqnarray*}
[\xi_{(1)},\xi_{(3)}]=-\xi_{(2)},\quad
[\xi_{(2)},\xi_{(3)}]=\xi_{(1)},\quad
[\xi_{(4)},\xi_{(5)}]=-\sqrt{\psi}\xi_{(6)},
\end{eqnarray*}
\begin{eqnarray*}
[\xi_{(4)},\xi_{(6)}]=\sqrt{\psi}\xi_{(5)},\quad
[\xi_{(5)},\xi_{(6)}]=-\frac{1}{\sqrt{\psi}}\{\frac{\psi}{T_{1}}
-(\frac{\dot{T}_{1}}{2T_{1}\sqrt{T_{0}}})^2\}\xi_{(4)},
\end{eqnarray*}
\begin{eqnarray*}
[\xi_{(i)},\xi_{(j)}]=0,\quad\mathrm{(otherwise)}.
\end{eqnarray*}
The \textbf{case 2iib$(**$)} leads to the four MCs out of which
three are usual plane symmetry KVs and one is proper MC.

\section{Examples}

In this section, we shall present some examples which satisfy the
constraints.\\
\par\noindent
1. The following metric satisfy the constraints of the case 1(i)
\begin{equation}
ds^2=e^{t^a}(dt^2-dx^2-dy^2-dz^2)
\end{equation}
which has six independent MCs.\\
\par\noindent
2. The solution of the constraint of the case 2ia$**(+)$ turns out
to be
\begin{eqnarray}
ds^2&=&dt^2-e^{\frac{2t}{a}}dx^2-e^{\frac{2t}{b}}(dy^2+dz^2),
\end{eqnarray}
where $a$ and $b$ are non-zero constants such that $a\neq b$. This
admits five independent MCs.\\
\par\noindent
3. When we solve the constraints of the case 2ia$**(++)$, we obtain
the following solution
\begin{equation}
ds^2=dt^2-dx^2-e^{\frac{2t}{a}}(dy^2+dz^2),\quad(a\neq0)
\end{equation}
which admits seven independent MCs.\\
\par\noindent
4. The metrics given below satisfy the constraints of the case
2ii(a)
\begin{eqnarray}
ds^2&=&t^2(dt^2-dx^2-dy^2-dz^2),\\
ds^2&=&dt^2-e^{\frac{2t}{a}}(dx^2+dy^2+dz^2),\\
ds^2&=&dt^2-(\frac{t}{t_0})^{2a}(dx^2+dy^2+dz^2)\quad(a\neq0)
\end{eqnarray}
which have ten independent MCs.

\section{Discussion and Conclusion}

In a recent paper [14], a classification of static plane symmetric
spacetimes has been obtained according to their energy-momentum
tensor. We have extended this classification to non-static plane
symmetric spacetimes satisfying $T_{01}=0$ for the degenerate case
only [19]. It was found four, six and ten independent MCs out of
which four are the usual KVs and the rest are the proper MCs. This
paper has been focussed to extend the classification to the
non-degenerate case along with $T_{01}=0$. The results are
summarized in the form of table 1.

\vspace{0.4cm}

{\bf {\small Table 1}. }{\small MCs for the Non-Degenerate Case}

\vspace{0.1cm}

\begin{center}
\begin{tabular}{|l|l|l|}
\hline {\bf Cases} & {\bf MCs} & {\bf Constraints}
\\ \hline 1(i) & $6$ & $(\frac{\dot{T}_{2}}{T_{2}\sqrt{T_{0}}})^.\neq0,~
(\sqrt{\frac{T_{1}}{T_{2}}})^.=0$
\\ \hline 1(ii) & $4$ &
$(\frac{\dot{T}_{2}}{T_{2}\sqrt{T_{0}}})^.\neq0,~
(\sqrt{\frac{T_{1}}{T_{2}}})^.\neq 0$
\\ \hline 2ia$(*)$ & $4$ &
$(\frac{\dot{T}_{2}}{T_{2}\sqrt{T_{0}}})^.=0,~
\frac{\dot{T}_{2}}{T_{2}\sqrt{T_{0}}}\neq0,~(\frac{T_{2}}{T_{1}})^.\neq0,~
(\frac{\dot{T}_{1}}{T_{1}\sqrt{T_{0}}})^.\neq0$
\\ \hline 2ia$**(+)$ & $5$ &$
\begin{array}{c}
(\frac{\dot{T}_{2}}{T_{2}\sqrt{T_{0}}})^.=0,~
\frac{\dot{T}_{2}}{T_{2}\sqrt{T_{0}}}\neq0,~(\frac{T_{2}}{T_{1}})^.\neq0,\\
(\frac{\dot{T}_{1}}{T_{1}\sqrt{T_{0}}})^.=0,~
\frac{\dot{T}_{1}}{T_{1}\sqrt{T_{0}}}\neq0
\end{array}$
\\ \hline 2ia$**(++)$ & $7$ &
$
\begin{array}{c}
(\frac{\dot{T}_{2}}{T_{2}\sqrt{T_{0}}})^.=0,~
\frac{\dot{T}_{2}}{T_{2}\sqrt{T_{0}}}\neq0,~(\frac{T_{2}}{T_{1}})^.\neq0,\\
(\frac{\dot{T}_{1}}{T_{1}\sqrt{T_{0}}})^.=0,~
\frac{\dot{T}_{1}}{T_{1}\sqrt{T_{0}}}=0
\end{array}$
\\ \hline 2i(b) & $10$ &
$
\begin{array}{c}
(\frac{\dot{T}_{2}}{T_{2}\sqrt{T_{0}}})^.=0,~
\frac{\dot{T}_{2}}{T_{2}\sqrt{T_{0}}}\neq0,~(\frac{T_{2}}{T_{1}})^.=0,
\end{array}$
\\ \hline 2iia$(*)$ & $10$ &
$
\begin{array}{c}
(\frac{\dot{T}_{2}}{T_{2}\sqrt{T_{0}}})^.=0,~
\frac{\dot{T}_{2}}{T_{2}\sqrt{T_{0}}}=0,~(\frac{(\sqrt{T_{1}})^.}{\sqrt{T_{0}}})^.=0,\\
\frac{(\sqrt{T_{1}})^.}{\sqrt{T_{0}}}\neq0
\end{array}$
\\ \hline 2iia$(**)$ & $10$ &
$
\begin{array}{c}
(\frac{\dot{T}_{2}}{T_{2}\sqrt{T_{0}}})^.=0,~
\frac{\dot{T}_{2}}{T_{2}\sqrt{T_{0}}}=0,~(\frac{(\sqrt{T_{1}})^.}{\sqrt{T_{0}}})^.=0,\\
\frac{(\sqrt{T_{1}})^.}{\sqrt{T_{0}}}=0
\end{array}$
\\ \hline 2iib$(*)$ & $6$ &
$
\begin{array}{c}
(\frac{\dot{T}_{2}}{T_{2}\sqrt{T_{0}}})^.=0,~
\frac{\dot{T}_{2}}{T_{2}\sqrt{T_{0}}}=0,~(\frac{(\sqrt{T_{1}})^.}{\sqrt{T_{0}}})^.\neq0,\\
(\frac{T_{1}}{2T_{0}}(\frac{\dot{T}_{1}}{T_{1}\sqrt{T_{0}}})^.)^.=0
\end{array}$
\\ \hline 2iib$(**)$ & $4$ &$
\begin{array}{c}
(\frac{\dot{T}_{2}}{T_{2}\sqrt{T_{0}}})^.=0,~
\frac{\dot{T}_{2}}{T_{2}\sqrt{T_{0}}}=0,~(\frac{(\sqrt{T_{1}})^.}{\sqrt{T_{0}}})^.\neq0,\\
(\frac{T_{1}}{2T_{0}}(\frac{\dot{T}_{1}}{T_{1}\sqrt{T_{0}}})^.)^.\neq0
\end{array}$
\\ \hline
\end{tabular}
\end{center}
We have obtained four, five, six, seven and ten independent MCs out
of which three are the usual plane symmetry KVs and the rest are the
proper. Further, it is mentioned here that each case has different
constraints on the energy-momentum tensor. Solving these constraints
may lead to some interesting solutions to the EFEs. We have found
some examples satisfying the constraints in some cases which help to
find the exact number of proper MCs. It would be interesting to
classify plane symmetric spacetimes according to their MCs for the
degenerate and non-degenerate cases by removing the assumption
$T_{01}=0$. This would lead to provide a complete classification of
non-static plane symmetric spaces according to their MCs.

\vspace{0.5cm}

{\bf \large References}

\begin{description}

\item{[1]} Katzin, G.H., Levine, J. and Davis, H.R.: J. Math. Phys.{\bf 10}(1969)617.

\item{[2]} Bokhari, A.H., Kashif, A.R. and Qadir, A.: J. Math. Phys. {\bf
41}(2000)2167.

\item{[3]} Hall, G.S. and da Costa, J.: J. Math. Phys. {\bf 32}(1991)2848; ibid 2854.

\item{[4]} Hall, G.S.: Gen. Rel. and Grav. {\bf 30}(1998)1099;
\emph{Symmetries and Curvature Structure in General Relativity}
(World Scientific, 2004).

\item{[5]} Hall, G.S., Roy, I. and Vaz, L.R.: Gen. Rel. and Grav. {\bf 28}(1996)299.

\item{[6]} Carot, J., da Costa, J. and Vaz, E.G.L.R.: J. Math. Phys. {\bf
35}(1994)4832.

\item{[7]} Carot, J. and da Costa, J.: {\it Procs. of the 6th Canadian Conf. on
General Relativity and Relativistic Astrophysics}, Fields Inst.
Commun. 15, Amer. Math. Soc. WC Providence, RI(1997)179.

\item{[8]} Tsamparlis, M., and Apostolopoulos, P.S.: J. Math.
Phys. {\bf 41}(2000)7543.

\item{[9]} Sharif, M.: Nuovo Cimento {\bf B116}(2001)673;
Astrophys. Space Sci. {\bf 278}(2001)447.

\item{[10]} Camc{\i}, U. and Barnes, A.: Class. Quantum Grav. {\bf
19}(2002)393.

\item{[11]} Camc{\i}, U. and Sharif, M.: Gen. Rel. and Grav. {\bf
35}(2003)97.

\item{[12]} Camc{\i}, U. and Sharif, M.: Class. Quantum Grav. {\bf 20}(2003)2169.

\item{[13]} Sharif, M. and Aziz, S.: Gen. Rel. and Grav. {\bf
35}(2003)1091;\\ Sharif, M.: J. Math. Phys. {\bf 44}(2003)5141.

\item{[14]} Sharif, M.: J. Math. Phys. {\bf45}(2004)1518.

\item{[15]} Sharif, M.: J. Math. Phys. {\bf 45}(2004)1532.

\item{[16]} Sharif, M.: Int. J. Mod. Phys. \textbf{D14}(2005)1675.

\item{[17]} Sharif, M.: Int. J. Mod. Phys. \textbf{A21}(2006)2645.

\item{[18]} Sharif, M. and Ismaeel, T.: Commun. Theore. Phys. \textbf{47}(2007)829.

\item{[19]} Sharif, M. and Ismaeel, T.: Mod. Phys. Lett. \textbf{A22}(2007)1813.

\item{[20]} Sharif, M.: Acta Physica Polonica \textbf{B38}(2007)2003.

\item{[21]} Stephani, H., Kramer, D., MacCallum, M.A.H.,
Hoenselaers, C. and Hearlt, E.: {\it Exact Solutions of Einstein's
Field Equations} (Cambridge University Press, 2003).

\end{description}

\end{document}